\documentclass[conference]{IEEEtran}
\IEEEoverridecommandlockouts
\usepackage{cite}
\usepackage{amsmath,amssymb,amsfonts}
\usepackage{algorithmic}
\usepackage{graphicx}
\usepackage{textcomp}
\usepackage{xcolor}
\usepackage[latin1]{inputenc}
\usepackage{graphicx}
\usepackage{epstopdf}
\usepackage{listings}
\usepackage{color}
\definecolor{keywordcolor}{rgb}{0.8,0.1,0.5}
\definecolor{webgreen}{rgb}{0,.5,0}
\usepackage[nice]{nicefrac}
\usepackage{epsfig}
\usepackage{graphicx}
\graphicspath{{./figs/}}
\usepackage{booktabs}
\usepackage{amsmath}
\usepackage{amssymb}
\usepackage{datetime}
\usepackage{bbm}
\usepackage{wasysym}
\usepackage{balance}
\usepackage{color,marvosym,xspace}
\usepackage{hyperref}
\usepackage{url}
\usepackage{multirow}
\usepackage{listings}
\usepackage{diagbox}
\usepackage{units}
\usepackage{subfigure}
\usepackage{enumitem}
\usepackage{ifsym}
\usepackage{lettrine}

\usepackage{relsize}
\usepackage{tikz}
\usepackage{pifont}
\usepackage{comment}

\def\BibTeX{{\rm B\kern-.05em{\sc i\kern-.025em b}\kern-.08em
    T\kern-.1667em\lower.7ex\hbox{E}\kern-.125emX}}
\begin{document}

\title{Sensing the Chinese Diaspora: How Mobile Apps Can Provide Insights into Global Migration Flows}

\author{\IEEEauthorblockN{Minhui Xue\IEEEauthorrefmark{1},
Xin Yuan\IEEEauthorrefmark{2}, Heather Lee\IEEEauthorrefmark{3}, and
Keith W. Ross\IEEEauthorrefmark{3}\IEEEauthorrefmark{4}}
\IEEEauthorblockA{\IEEEauthorrefmark{1} 
The University of Adelaide, Australia\\
\IEEEauthorrefmark{2} Beijing University of Posts and Telecommunications, China, and University of Technology Sydney, Australia\\
\IEEEauthorrefmark{3} New York University Shanghai, China\\
\IEEEauthorrefmark{4} New York University, USA}}

\maketitle

\begin{abstract}
Many countries today have ``country-centric mobile apps'' which are mobile apps that are primarily used by residents of a specific country. Many of these country-centric apps also include a location-based service which takes advantage of the smartphone's API access to the smartphone's current GPS location. In this paper, we investigate how such country-centric apps with location-based services can be employed to study the diaspora associated with ethnic and cultural groups. Our methodology combines GPS hacking, automated task tools for mobile phones, and OCR to generate migration statistics for diaspora.  As a case study, we apply our methodology to WeChat, an enormously popular app within China and among ethnic Chinese worldwide. Using WeChat, we collect data about the Chinese diaspora in 32 cities. 
The combined data provides interesting insights to the modern Chinese diaspora and how it has changed in recent years. 
\end{abstract}


\section{Introduction}

Diaspora studies lie at the intersection of the humanities and social sciences, encompassing disciplines as diverse as geography, sociology, history, anthropology, psychology, political science, and statistics. Data for diaspora analysis has traditionally come from surveys and censuses. These approaches, however, tend to lack quantitative up-to-date information and are subject to significant biases resulting from undercoverage and low response rates. Snowball and convenience sampling -- as two popular sample surveys -- are costly and time consuming, and are typically limited to a small number of cities or regions~\cite{belai2007enabling}. Mail-out surveys must be tailored to each respondent's preferences and interests, rendering response rates stubbornly low~\cite{bloch2005development,nworah2005study}. 

In this paper, we propose using ``country-centric mobile apps'' for collecting fine-grained geographic and time data for worldwide diaspora. Country-centric mobile apps are mobile apps that are primarily used by residents of a specific country.  For example, the immensely popular WeChat app is used by more than 1.1 billion people, with the large majority of these users being of Chinese heritage, including Chinese citizens living in China or ethnic Chinese in the worldwide Chinese diaspora. In China today, essentially {\em all} messaging is done over WeChat, as SMS and other messaging services have been fully supplanted by WeChat. Similarly, in South Korea the vast majority of the messaging takes over the country-centric app Kakao Talk, and in Russia most of the messaging takes place over VK. Yik Yak, an anonymous message board app, is primarily used by students in U.S. universities. Table~\ref{tab:nearby} lists some popular country-centric and culture-centric mobile apps.

 \begin{table}[b]\footnotesize
\vspace{-0.4cm}
 \caption{Country-Centric and Culture-Centric Mobile Apps}
 \vspace{-0.1cm}
 \label{tab:nearby}
 \begin{minipage}{\columnwidth}
 \begin{center}
  {\small\resizebox{\textwidth}{!}{
 \begin{tabular}{c|c|c}
\toprule
{\bf Mobile App} & {\bf Number of Users Worldwide}  & {\bf Primary Country/Culture}   \\ 
\midrule
WeChat &1.1 billion & China \\
Tantan &2.5 million & China \\
LINE & 560 million& Japan \\
Kakao Talk &170 million& South Korea \\
VK &370 million& Russia \\
Zalo &30 million& Vietnam \\
imo &150 million& Cuba \\
Yik Yak &4 million (monthly)& USA\\
JDate & 0.45 million & Jewish \\
\bottomrule
\end{tabular}}}
\end{center}
\end{minipage}
\end{table}

Many of these country-centric mobile apps also include a location-based service which takes advantage of the smartphone's API access to the smartphone's current GPS location. For example, the Yik Yak application displays bulletin board messages that are within a few kilometers of the user. The WeChat application includes a People-Nearby service which displays WeChat users who are within a few kilometers. In fact, most of the apps shown in Table~\ref{tab:nearby} provide a location-based service similar to WeChat's People-Nearby. 

In this paper, we will show how country-centric apps with location-based services can be used to collect fine-grained geographic data and time-series about diaspora. Although the methodology in this paper can potentially applied to a variety of ethnic and cultural groups (Kakao Talk for the Korean diaspora, VK for the Russian diaspora, JDate for the Jewish diaspora, and so on), we will focus on using WeChat to study the Chinese diaspora. 

The contributions of this paper are as follows:
\begin{itemize}[wide=1pt,leftmargin=10pt]
\item We develop an automated data collection methodology for collecting usage and social media information from country-centric mobile apps. The method can be scaled to hundreds of cities, and does not require the mobile app to provide an HTML interface that can be scrapped. Although we develop the methodology for WeChat, it can be applied to other country-centric mobile apps.
\item Applying this methodology to WeChat's People Nearby service, we take measurements for 32 cities on four consecutive Saturday afternoons. 
\item We show how the WeChat People Nearby data can be mined to obtain relative estimates of the Chinese diaspora among the cities. As the WeChat People Nearby service is primarily used by users in the 18 to 30 years old demographic, we argue that the relative estimates provide insight into immigration movements for younger generations. 
\item As there is no ground truth, we corroborate the estimates by providing journalistic evidence of recent Chinese growth, which matches what is being observed from our mobile data. 
\item By investigating the social media posts, we also investigate the extent WeChat users are employing the receiving country's native language. High rates of employment implies a high degree of assimilation. 
\end{itemize}

\section{Related Work}\label{related_work}

Surveys have been traditionally used to gather data about diaspora (Chinese and otherwise). Sources have included census, immigration data (particularly from destination countries), and sample surveys. These approaches, however, tend to lack quantitative up-to-date information and are subject to significant biases resulting from undercoverage and low response rates.  
The survey methodologies are ``active'' where the respondents must actively participate in the survey. However, more recently, there has been interest in developing ``passive'' Internet-based methodologies which can provide contextually-rich data more readily and more broadly than the traditional active methodologies. 

Specifically, Zagheni and Weber~\cite{zagheni2012you} and State \textit{et al.}~\cite{weber2013studying} estimated international migration rates using IP-geolocated data of millions of anonymized Yahoo users' logins. 
One drawback of this approach is that it does not allow for a fine-grained geographic analysis as IP addresses have low granularity within a country: they cover large regions sometimes entire countries and are not correct all the time. Zagheni \textit{et al.}~\cite{zagheni2014inferring} and Hawelka \textit{et al.}~\cite{hawelka2014geo} have used geo-located Twitter tweet data to estimate international migration rates and trends. 
Although geolocated Twitter data can provide geographical information and the scope is global, these geo-located tweets only represent a tiny fraction of the Twitter tweets, and their estimates of geographic coordinates can also be manually edited by users, which inevitably causes errors. 
Most recently, Messias \textit{et al.}~\cite{messias2016migration} used  ``places lived'' from Google+ profiles, crawled in 2012, to study migration clusters of country triads rather than country-to-country migration flows. Google+ data does not provide the chronological order where people have lived, and Google+ has almost zero usage rates in many countries. 

To our knowledge, this is the first paper to employ country-centric mobile apps to study diaspora. As compared to the above Internet-based approaches, our approach based on mobile app location-based services (in this paper, WeChat's People-Nearby service) has several advantages:
\begin{itemize}[wide=1pt,leftmargin=10pt]
\item Mobile services such as People-Nearby and messaging services are used by a large swath of the target population, and are not limited to highly-educated users. 
\item Compared with IP-based approaches, GPS hacking can provide fine-grained geographical estimates.
\item Measurements can be taken not only from any specific location, but also at any specific time. Unlike the other active approaches, our approach  can provide time series information, estimating migration flows on a month-to-month basis. 
\item Compared with Twitter and Google+ based approaches, mobile app People-Nearby services can provide precise estimates of geographic coordinates by applying localization methodology~\cite{xue2015thwarting,ding2014stalking}. They can also provide real-time demographic information, though the time series are relatively short.
\end{itemize}

\section{Methodology}\label{method}

Although the methodology we develop in this paper can be used with many country-centric mobile apps that include location-based services, we will use WeChat's People Nearby service as a case study for the methodology. 

\subsection{WeChat's People Nearby Service}

WeChat's People Nearby service takes as input the current geo-location of a user's mobile device and returns a list of WeChat users in close proximity. A user can then send requests to people returned by the service, with the hope of being accepted as a WeChat friend. Once a friendship relationship is established between two users, the users can message each other and see each other's social media postings. 

For each user listed in People Nearby, WeChat gives an indication of how close the user is. Specifically, using bands of 100 meters, WeChat reports that a user is within 100 meters, or within 200 meters, or within 300 meters, and so on. For example, if both Alice and Bob are using the People Nearby service, and Alice is 368 meters away from Bob, then WeChat will report to Alice that Bob is within 400 meters of her. We emphasize that the People Nearby service only returns people who have recently used People Nearby; it does not return all WeChat users. 

  \begin{figure}[t]
  \centering
  \mbox{
    \subfigure[People-Nearby Button\label{fig:people_nearby_button}]{\includegraphics[scale = 0.14]{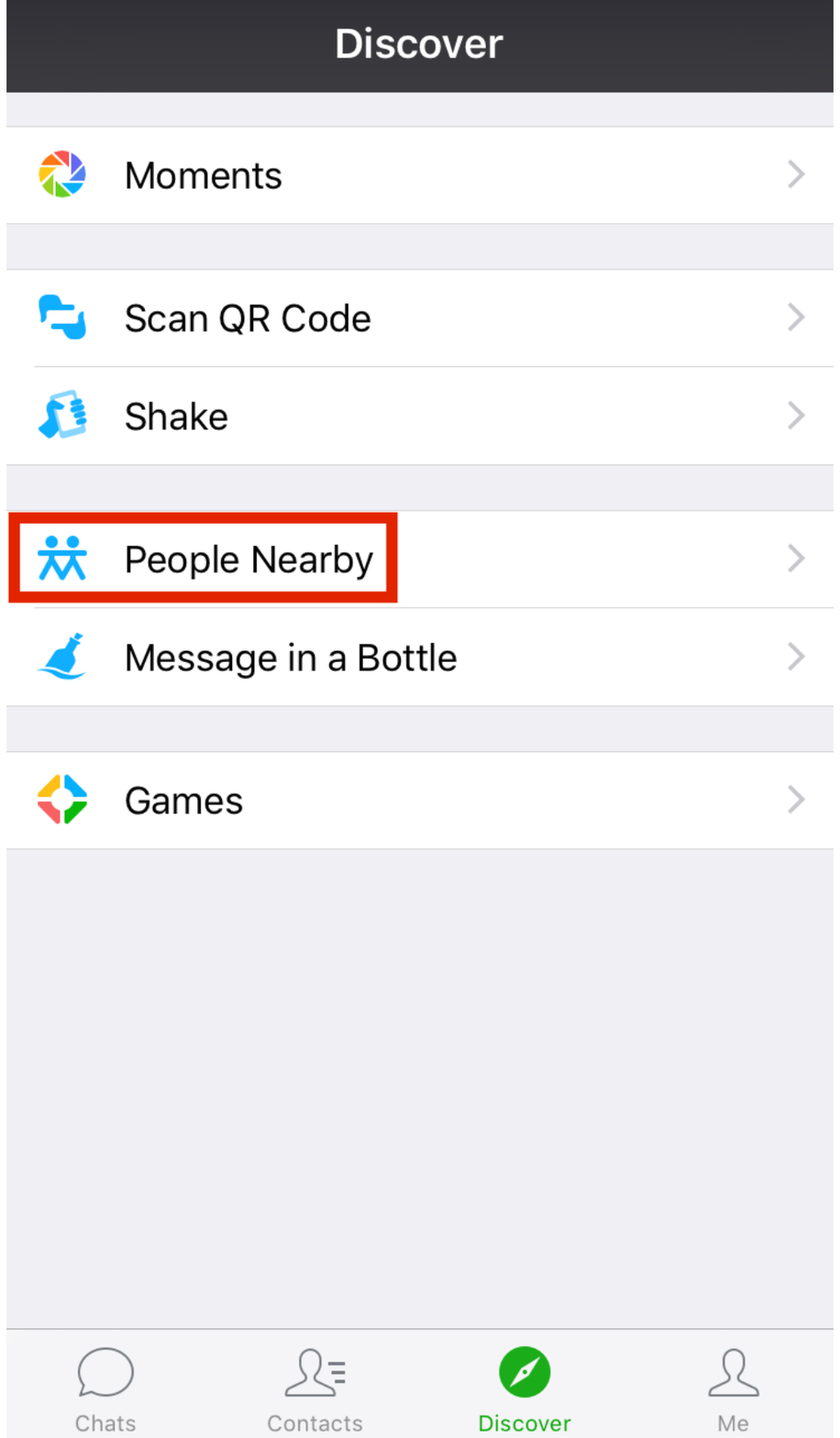}}
    \subfigure[People-Nearby List\label{fig:people_nearby_list}]{\includegraphics[scale = 0.105]{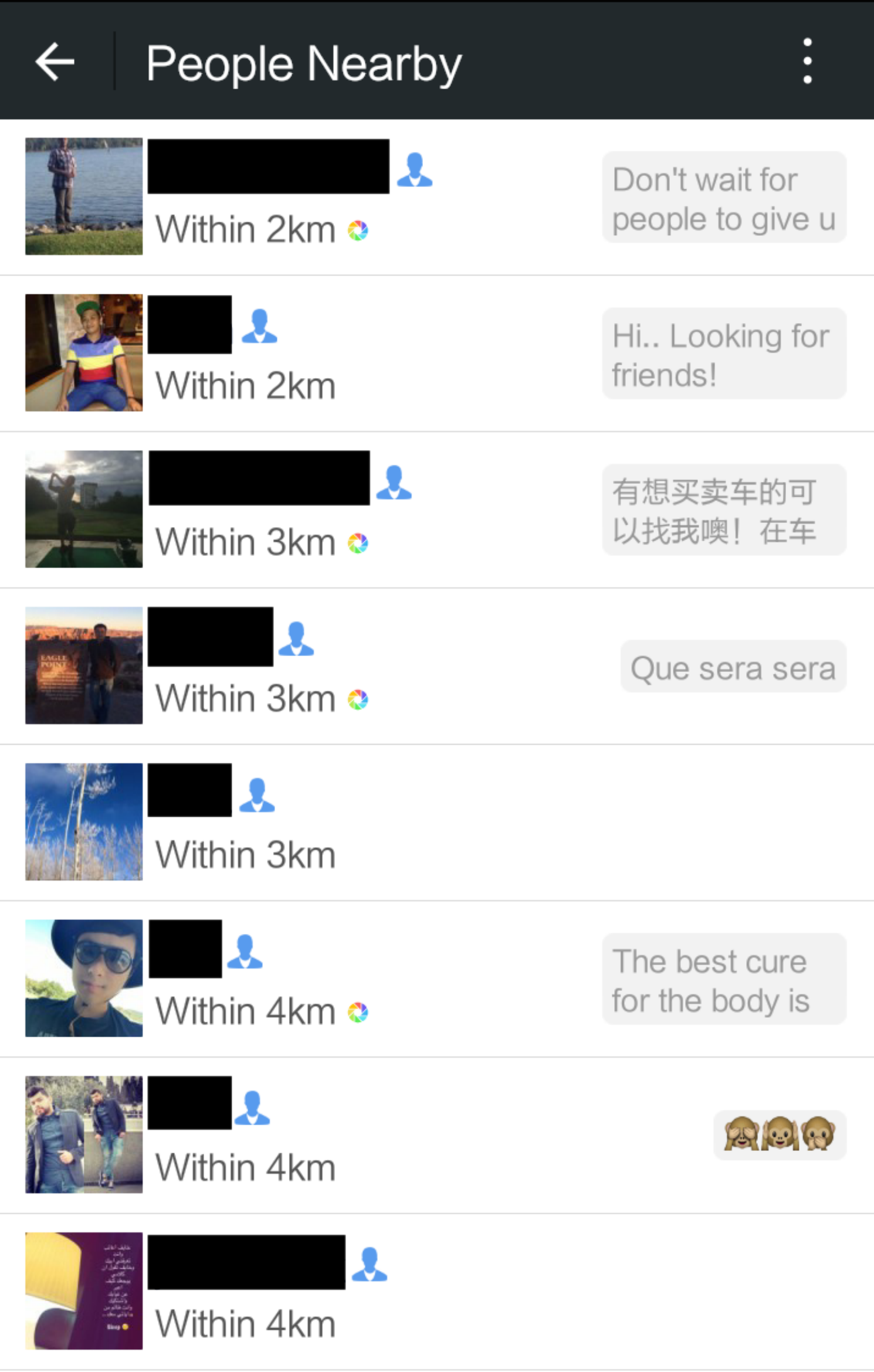}}
    \subfigure[Moments\label{fig:moment}]{\includegraphics[scale = 0.14]{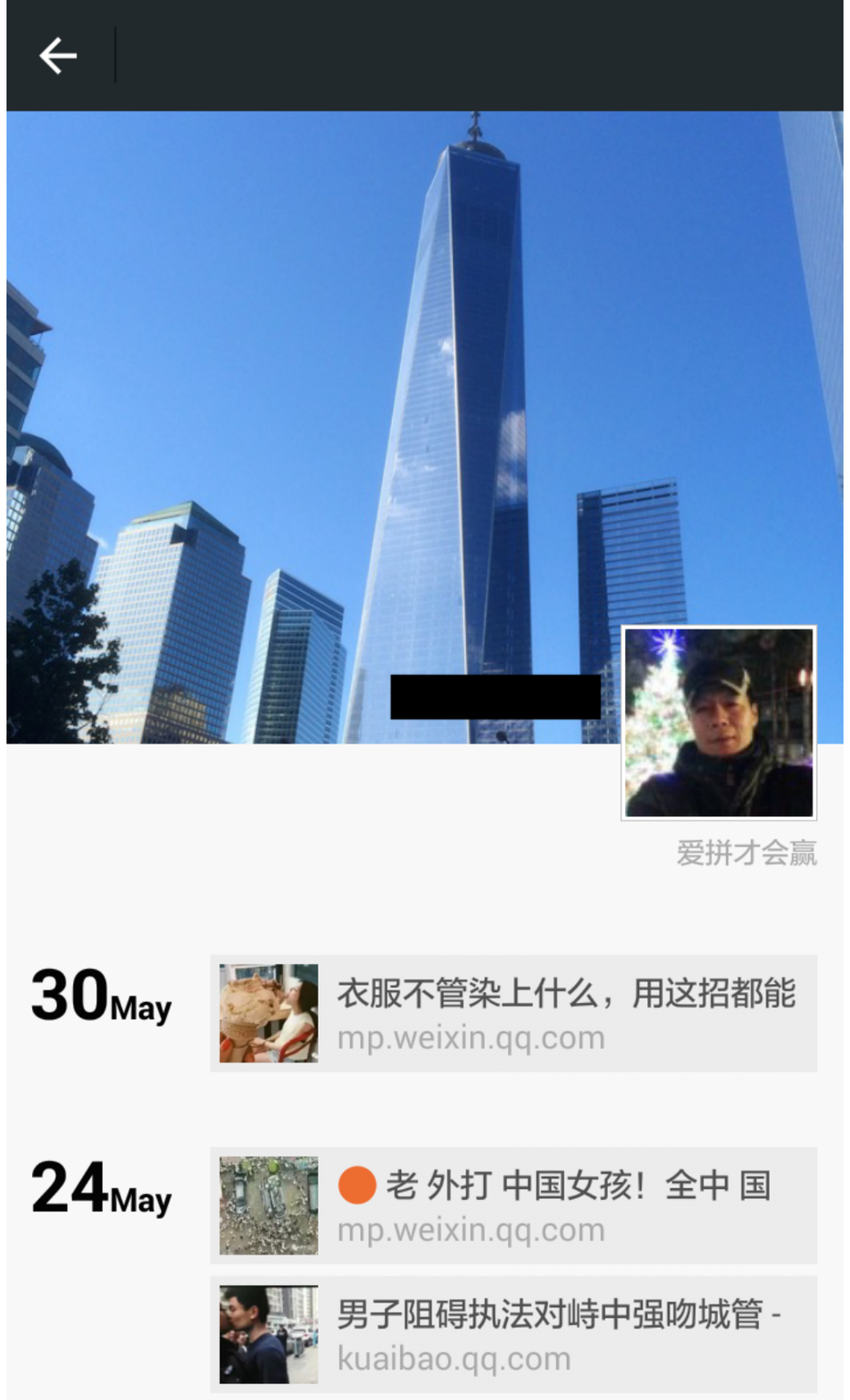}}
    }
    \vspace{-0.2cm}
  \caption{Functionality of the WeChat People-Nearby Service}
  \label{WeChat_People_Nearby_service}
  \vspace{-0.5cm}
\end{figure}

Figure~\ref{WeChat_People_Nearby_service} shows the functionality of the People-Nearby service. Figure~\ref{fig:people_nearby_button} shows the screen with the People-Nearby Button (among other buttons); Figure~\ref{fig:people_nearby_list} shows a typical list of people returned by People-Nearby. (In the figure, we have redacted the names for privacy purposes.) Note that WeChat lists the nearby people starting from nearest to farthest and provides the different distance ranges. When Alice taps on  a person in the list, say Bob, Alice will see Bob's profile page and will be able to send a greeting. After Bob receives the greeting, he can decide whether or not he wants to befriend Alice. The granularity of the distances that WeChat reports is non-linear; distances up to 1,000 meters are reported in bands of 100 meters, but beyond 1,000 meters, the band size increases to 1,000 meters increments. Figure~\ref{fig:moment} shows visible moments for a Chinese user.

For a given set of target cities, the WeChat People-Nearby service can provide an indication of relative sizes of ethnic populations in the cities. For example, suppose that we use the WeChat People-Nearby in Antwerp and Brussels at exactly the same time, and we find that 80 users are within 2,000m of the center of Brussels and 40 users are within 2,000m of the center of Antwerp, giving a ratio of 2 to 1. Suppose we further repeat this experiment on different days and observe similar ratios between Antwerp to Brussels users. Then we can estimate that the number of ethnic Chinese residents in Brussels is roughly twice that of Antwerp. 

These relative estimates are admittedly rough estimates and have inaccuracies for a number of reasons. The largest source of error is that the users displayed by People-Nearby are not a random sample of the ethnic Chinese residents in the city. Although WeChat today is used by most Chinese nationals between 15 and 40 and also by many people over 40, by inspecting the photos in our collected data we see that the People-Nearby service is predominantly employed by users in the 18-35 age group. Therefore, our methodology provides an indication of the relative populations of young (18 to 35 years old) ethnic Chinese, a demographic that should reflect recent immigration flows into the the cities over the past 20 years. 

Another factor contributing to inaccuracies is that not all WeChat users are ethnic Chinese. Indeed, as we will see in this study, in many cities a significant fraction of WeChat People-Nearby users are non-ethnic Chinese. However, the non-Chinese users can be separated from the ethnic Chinese users either manually or by automated procedures, as we will discuss in Section~\ref{Labeling_Chinese}.

We will argue that the methodology can indeed provide reasonably good comparisons of the number of young ethnic Chinese in various cities around the world. Moreover, we will see that the data obtained from WeChat People-Nearby can be compared with other datasets to gain deeper insights into the migration flow dynamics.

\subsection{Data Collection Methodology}\label{collection_methodology}

There are a number of challenges in collecting WeChat People Nearby data from a large number of cities:
\begin{itemize}[wide=1pt,leftmargin=10pt]
\item We need to run the WeChat People Nearby service in each of the target cities without actually being physically located in each of the target cities. Furthermore, in order to have a meaningful comparison, it should be run in each target city at the same local time. 
\item Because WeChat is a smartphone application, we cannot collect the data by simply scraping HTML pages. 
\item We need to develop an automated approach for collecting the data, as manually running the People Nearby service repeatedly in many cities is labor intensive. 
\end{itemize}

 \begin{figure}[t]
   \centering
   \includegraphics[scale = 0.28]{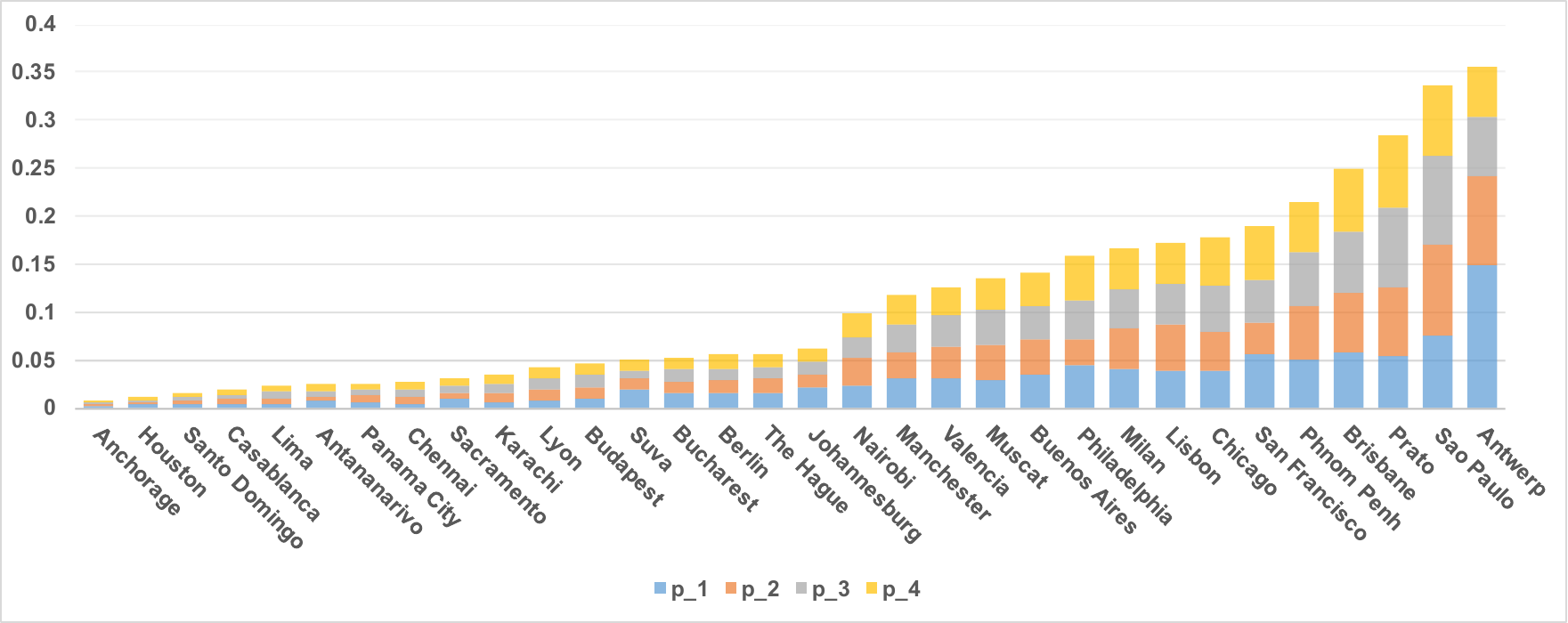}
   \vspace{-0.2cm}
   \caption{Consistency Across Experiments}
  \label{fig:consistency2000_1}
  \vspace{-0.3cm}
\end{figure}

In order to run People-Nearby in each of the cities, we employ GPS hacking to collect data from worldwide locations without having a physical presence in the locations. It is well known that a smartphone's GPS location can be faked so that applications running on the smartphone, and the corresponding backend servers, think that the smartphone is physically located in those locations. Faking a smartphone's location in such a manner is referred to as {\em GPS hacking}. For example, with GPS hacking, a user physically located in Paris can set its smartphone GPS coordinates to a specific street corner in Brooklyn, and then use Yelp to browse restaurant reviews in Brooklyn, or use Waze to see current traffic conditions in Brooklyn. For the data collected in this paper, we use smartphones physically located in Shanghai and set the GPS locations of the smartphones to be those of the city halls of the target cities. We did not collect more data so as to not overly burden the WeChat servers. 

Because WeChat is a smartphone application, we cannot collect the data by simply scraping HTML pages. To obtain the People-Nearby data, it may be possible to reverse engineer the protocol or employ a man-in-the-middle attack, although this seems difficult with recent versions of WeChat~\cite{wang2016defending}. Instead, we develop a more generic methodology for obtaining the data, one that should work for virtually all smartphone applications. This methodology has two components: (1) a task automation tool, which automatically sets the GPS location, taps on the various WeChat People-Nearby buttons, scrolls to the bottom of the People-Nearby page, and takes screenshots of the screens displayed by People-Nearby and Moments; (2) an optical character recognition (OCR) tool that extracts the textual data (in Chinese and Latin characters) from the screenshots. For the task automation tool, we use Appium\footnote{\url{http://appium.io/}} to execute scripts in a smartphone running WeChat. For each city and each Saturday, it takes 20-40 minutes to collect all the relevant screenshots. After collecting all the screenshots, we  process them by using the optical character recognition tool Tesseract-OCR,\footnote{\url{https://github.com/tesseract-ocr}} which can recognize both Chinese and Latin characters. We extract usernames, reported distances, status messages, and social networking postings for each discovered user in the target city.

\subsection{Labeling the Users as Chinese}\label{Labeling_Chinese}

Depending on the city, a significant fraction of the users listed by People-Nearby are not ethnic Chinese. As we are interested in using WeChat to study the Chinese diaspora, we need to filter out the non-ethnic Chinese. To this end, we tried two approaches, a manual approach and an automated approach. For the manual approach, we asked three Chinese undergraduate students to label each collected WeChat user; for the rare cases where there was not a consensus, we used voting. To carry out the labeling, the students considered the profile picture and whether the username, the status, or information in postings contain Chinese characters and/or pinyin words.

The automated classified users as Chinese if they employed Chinese characters in their posts. We found this technique to have 76\% accuracy, with errors primarily due to some Chinese users writing in English or in the local native language. To improve the accuracy of the automated approach, we would need to  not only examine the characters in the posts, but also classify the profile photos as ethnic-Chinese or not. Although we believe this is feasible, it is a major research project on its own and is beyond the scope of this paper. Throughout this paper, the results reported in this paper use the manual labeling of Chinese and non-Chinese users.

\section{The Dataset}\label{data}
For this study, we selected 32 cities from Africa, Asia, Europe, North America, and South America. In choosing cities, we tried to cover most time zones and also cover both large and small cities. The cities greatly differ in population, ranging from about 20,000 to 12 million residents. The list of cities can be seen in Figure~\ref{fig:consistency2000_1}. We used the automated procedure described in Section~\ref{collection_methodology} to collect the data on four Saturdays at 3:00 PM local time for each city, from 18 June 2016 to 9 July 2016. In total, the automated procedure generated over 75,000 screenshots. In order to have a fair comparison among the cities, we only consider users reported within 2,000 meters of each of the city halls.\footnote{A more thorough study would take measurements from different locations within large cities. In this proof of concept, we only use one location for each city.} From the screenshots, for users within 2,000 meters in each city, we obtained profile and social media information for 6,308 distinct users, of which 3,205 were labeled as ethnic Chinese.

Let $x_j$ and $y_j$ be, across the four experiments, the total number of distinct Chinese WeChat users and the total number of distinct WeChat users within 2,000 meters, respectively. Table~\ref{tab:dataset} shows some basic statistics for the dataset. 

 \begin{table}[t]\footnotesize
 \caption{High-level Statistics for WeChat Users Across Cities}
\vspace{-0.2cm}
 \label{tab:dataset}
 \begin{minipage}{\columnwidth}
 \begin{center}
  {\small\resizebox{\textwidth}{!}{
 \begin{tabular}{c|c|c}
 \toprule
{} & {\bf \# Distinct WeChat Users}  & {\bf \# Distinct Chinese}\\ 
{} & {}  & {\bf  WeChat Users}\\ 
\midrule
Maximum (in city) & 602 (Sao Paulo) & 287 (Prato)\\
Minimum (in city) & 7 (Anchorage)& 7 (Anchorage)\\
Average (across cities)& 197.1 & 100.2\\
Median (across cities)& 173.5 & 52\\
\bottomrule
\end{tabular}}}
\end{center}
\end{minipage}
\vspace{-0.2cm}
\end{table}

\subsection{Data Consistency over Time}

We now investigate whether the relative number of People-Nearby users in different cities changes significantly from one week to another. If the relative number of users is more or less constant, then the four snapshots taken on each of the four Saturdays should be representative of the relative People-Nearby usage across cities. 

Let $x_{ij}$  and $y_{ij}$ be the total number of Chinese WeChat users and distinct WeChat users observed within 2,000 meters of the city hall in city $j$ during experiment $i$. To investigate the degree of data consistency, let
\begin{equation}
p_{ij} = \frac{x_{ij}}{\sum_{j=1}^{32} x_{ij}}
\end{equation}
be, among all the Chinese users observed in experiment $i$ across all cities, the proportion of observed Chinese users in city $j$ in experiment $i$.  Ideally these proportions should be more or less constant in each of the four experiments. Figure~\ref{fig:consistency2000_1} shows for, each of the 32 cities, the four values of $p_{ij}$, where each proportion $p_{ij}$, $i=1,\dots,4$ is shown as a vertically aligned bar.

From Figure~\ref{fig:consistency2000_1}, we see the four vertical bars for each city are similar in size and that they are consistently increasing from city to city. This serves to show that the proportion of Chinese users in each city remains relatively constant across experiments. However, because there is some minor statistical variation, we will aggregate over the four experiments when presenting the results in the subsequent section.

\section{Data Analysis}\label{data_analysis}

\begin{figure}[t]
   \centering
   \includegraphics[scale = 0.28]{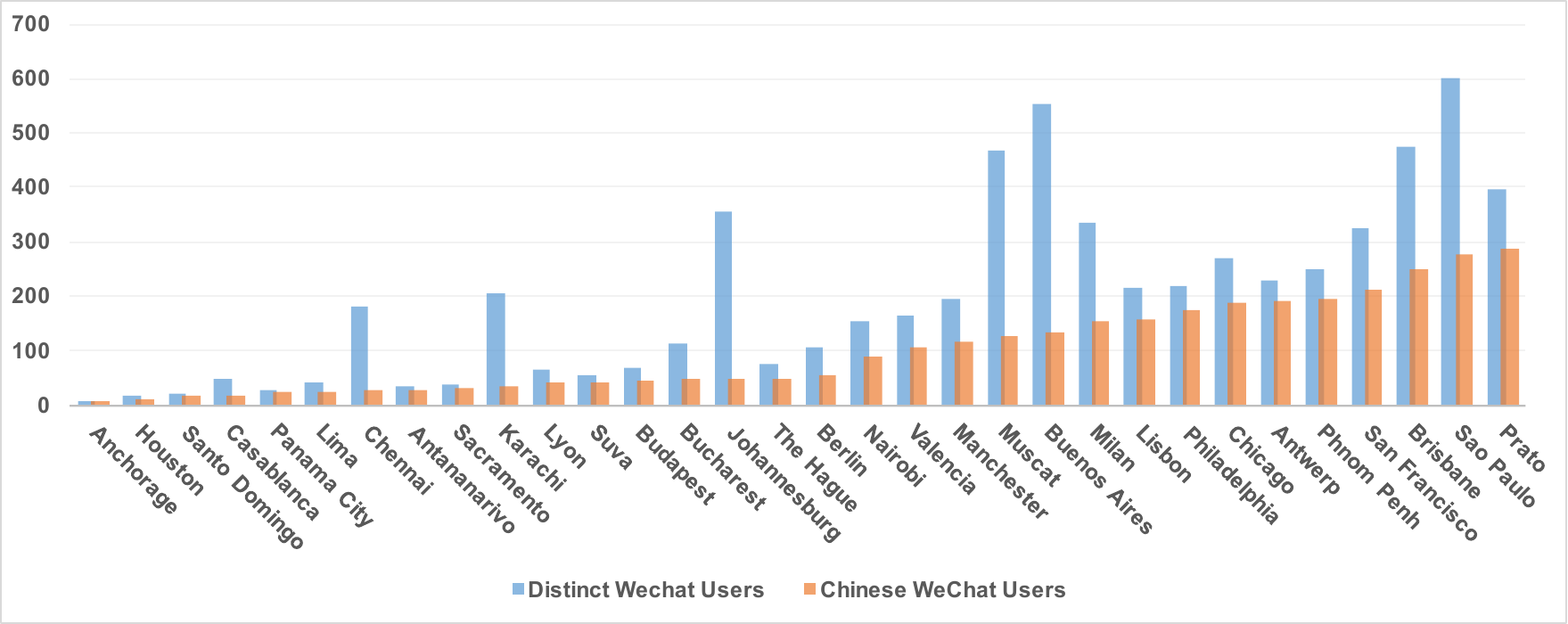}
   \vspace{-0.2cm}
   \caption{Number of Distinct WeChat Users versus Chinese Users}
   \label{fig:distinct2000}
   \vspace{-0.4cm}
\end{figure}

Figure~\ref{fig:distinct2000} shows the $x_j$ (total number of Chinese users observed in city $j$) and $y_j$ (total number of users observed in city $j$) values for the 32 cities, where the cities are ordered from lowest to highest $x_j$ value. We see that for most of the cities, the Chinese WeChat users represent more than 70\% of the WeChat users. Nevertheless, in Buenos Aires, Muscat, Johannesburg, Karachi, and Chennai, Chinese users are only a small fraction of the overall WeChat users. This is likely because WeChat is being aggressively marketed to the general population in Africa, India, and South America~\cite{fin24tech,techasia,Forbes}. (Nevertheless, in these same continents and countries, WeChat to date has had overall low penetration rates relative to other messaging apps such as Whatsapp, Messager, Viber, and Line~\cite{techasia2,similarweb}.) We also see in Figure~\ref{fig:distinct2000} that among the cities considered, the cities with the most Chinese People-Nearby users, and thus likely with the largest population of young ethnic Chinese, are Prato (Italy), Sao Paulo (Brazil), Brisbane (Australia), San Francisco (USA). These cities are then followed by Phnom Penh (Cambodia), Antwerp (Belgium), Chicago (USA), Philadelphia (USA), and Lisbon (Portugal). 

Let $s_j$ be the population for city $j$.\footnote{Ideally, we would like $s_j$ to be the population for within 2,000 meters of city hall for city $j$. But this data is difficult to acquire for many countries.} We define $x_j/s_j$ and $y_j/s_j$ to be the  concentration of Chinese People-Nearby users and the concentration of People-Nearby users in city $j$, respectively.  (The values in Figure~\ref{fig:population} are shown on a log scale and are further normalized so that the city with the smallest value of  $x_j/s_j$ is set to~$1$.) In Figure~\ref{fig:population} we order the cities from lowest to highest values in terms of the concentration of the Chinese users. We see that Prato (Italy) has by far the highest concentration of People-Nearby users, with a concentration of more than twice that of the second most concentrated city. Following Prato, the cities with the highest concentrations are Suva (Fiji), Antwerp (Belgium), Lisbon (Portugal), and San Francisco (USA).

\subsection{Cities with Highest Concentration of Chinese People Nearby Users}

\begin{figure}[t]
   \centering
   \includegraphics[scale = 0.28]{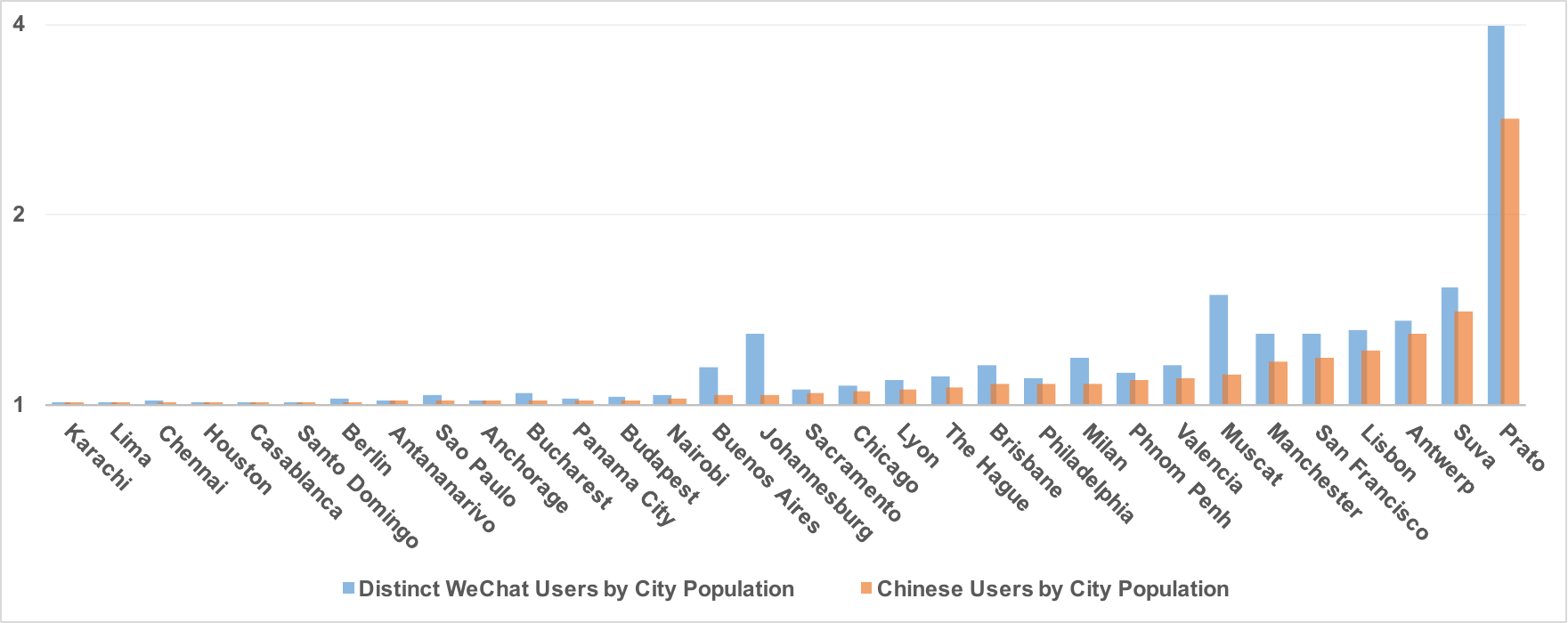}
  \vspace{-0.2cm}
   \caption{Number of WeChat Users and Chinese Users Normalized by City Population}
   \label{fig:population}
   \vspace{-0.3cm}
\end{figure}

Prato, 20km from Florence, is renowned for its historic textile industry. Prato is also the home to one of the largest populations of Chinese residents in Europe, a phenomenon that is remarkable not only for its magnitude but also for the speed with which it has developed~\cite{PratoBook}. The first Chinese people came to Prato in the early 1990s. The majority work in 3,500 workshops in the ready-to-wear garment industry. According to Wikipedia, Prato has the largest concentration of Chinese people in all of Europe~\cite{WikiItaly}. Given that Prato has the largest concentration of ethnic Chinese in Europe, and that most of the Chinese people immigrated to Prato in the past 25 years, we would expect it to have one of the highest density of People-Nearby users. Indeed, among the 11 European cities considered in this study, as shown in Figure~\ref{fig:population}, it has by far the highest concentration of Chinese People-Nearby users, demonstrating that our People-Nearby methodology can be used to track recent Chinese labor migrant communities. 

Suva is the capital of Fiji, an island country in the South Pacific Ocean. Although up until recently Suva had a relatively small Chinese population (approximately 1\% of the residents are ethnic Chinese), in recent years China has  been a big source of investment and tourists in Fiji~\cite{Pirepoint}. The recent Chinese business and tourism activity in Fiji explains why Fiji is also among one of the cities with the highest concentration of Chinese People WeChat users. 

Antwerp, a famous diamond city in Belgium, has a Chinatown, whose beginning dates back to the 1970s as a result of Chinese migration after the World War II~\cite{AntwerpChinatown}. Lisbon is the home of Portugal's largest Chinatown. Up until the recent past, most Chinese immigrants to Portugal came from the former Portuguese colony, Macau. Antwerp and Lisbon are among  the cities with the highest concentration of Chinese People-Nearby users. Although Lisbon has a Chinatown, it has not been known as a city with usually large concentrations of ethnic Chinese people. The results here may point to increased Chinese immigration to Lisbon in recent years~\cite{Portugal}.

San Francisco is well known to have a large ethnic Chinese population. According to the 2010 U.S. census, 21\% of the population in San Francisco was of Chinese descent with at least 150,000 Chinese American residents~\cite{2010UScensus}. San Francisco has the highest percentage of residents of Chinese descent of any major U.S. city, and the second largest Chinese American population, after New York City. Although San Francisco is among the cities with the highest concentrations of Chinese People-Nearby users, given that it has had a thriving Chinese community since the 19th century, it is perhaps surprising that the concentration of Chinese People-Nearby users is not even higher than reported in Figure~\ref{fig:population}. This is perhaps partially due to the fact that many of the ethnic Chinese living in San Francisco are over 40 and are not active People-Nearby users. It is also likely because we are only counting people found by People-Nearby within 2,000 meters of the city hall. If we had taken readings from a larger collection of probes scattered throughout the city, the concentration of People-Nearby users would likely be much higher.

\begin{figure}[t]
   \centering
    \subfigure[Chinese WeChat Users Language Usage (English is the Main Language)\label{fig:language_1}]{\includegraphics[scale = 0.28]{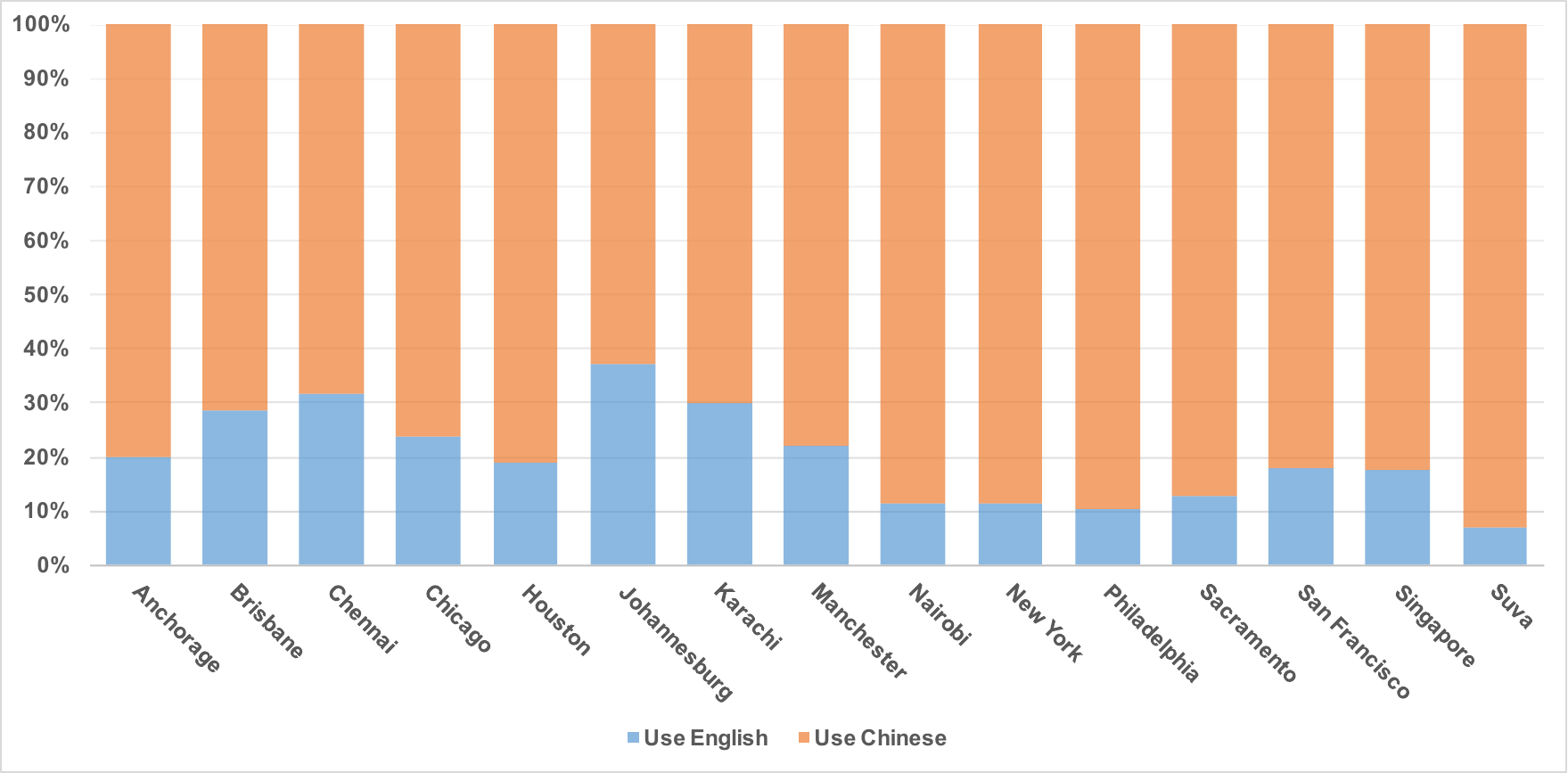}}
    \subfigure[Chinese WeChat Users Language Usage (English is Not the Main Language)\label{fig:language_2}]{\includegraphics[scale = 0.28]{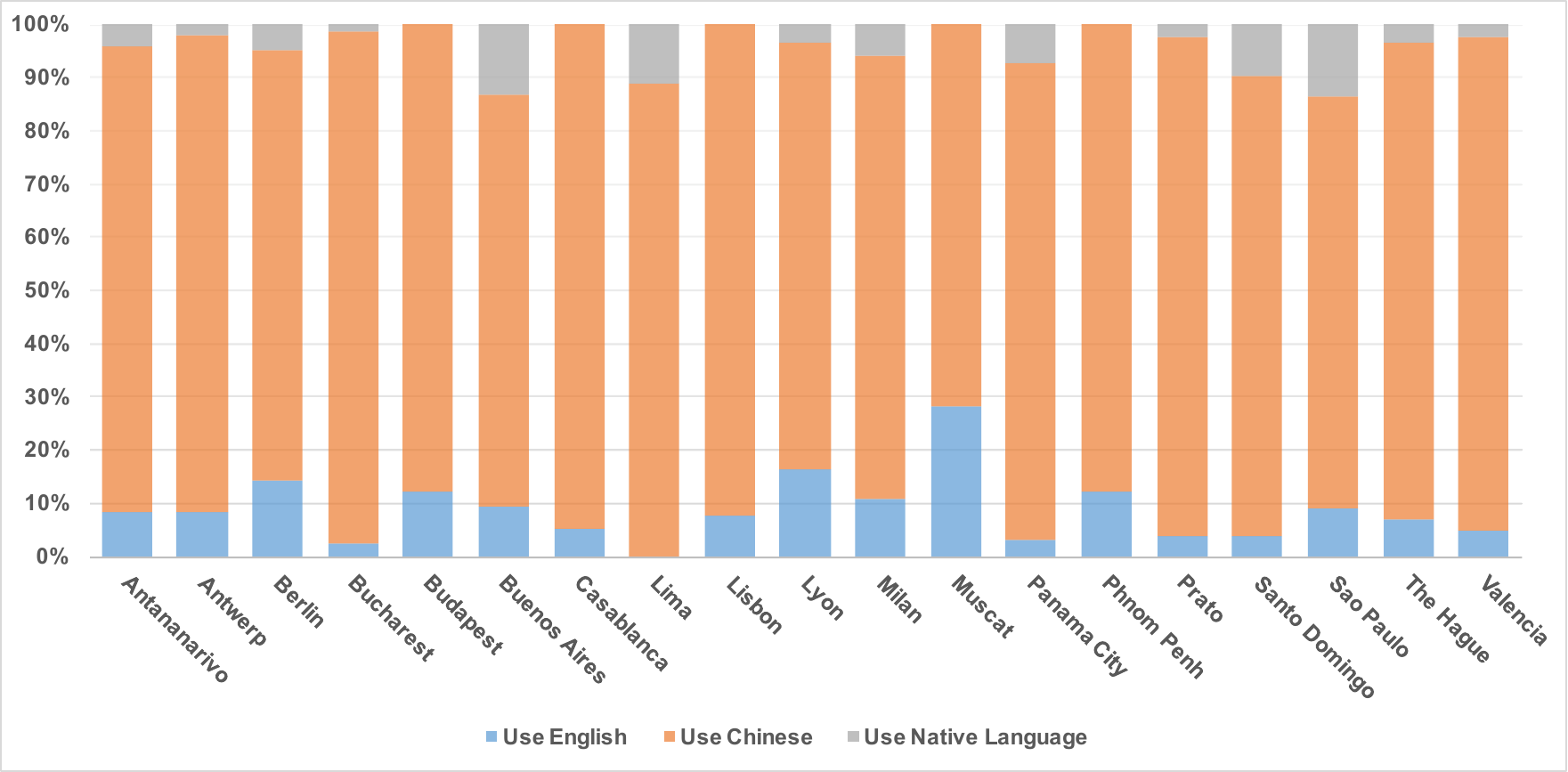}}
    \vspace{-0.2cm}
    \caption{Chinese WeChat Users Use of Local Language}
    \label{fig:language}
    \vspace{-0.4cm}
\end{figure}

\subsection{Language Assimilation}

An immigrant's assimilation into a receiving country can be measured to some degree by whether he or she is using the local language of the receiving country. By examining the text that is employed in the WeChat status messages and Moments, we can see whether the Chinese user is using the local language. If a user uses the local language at all (as well as Chinese), we consider that person to be a user of the local language. 

Figure~\ref{fig:language} shows the language usage distribution of Chinese WeChat users. As we see from Figure~\ref{fig:language}, a significant fraction of the Chinese WeChat users employ the local  language in Johannesburg, Chennai, Karachi, Muscat, Lyon, and Berlin. In many cities, such as Prato and Bucharest, there is very little usage of the local language or of English, indicating possibly low levels of assimilation. In these cities, such low levels of assimilation are likely due to education and labor markets~\cite{herdaugdelen2016social,gordon1964assimilation}.

\section{Conclusion}\label{conclusion}

Many countries today have ``country-centric mobile apps'' which are mobile apps that are primarily used by residents of a specific country. Many of these country-centric apps also include a location-based service which takes advantage of the smartphone's API access to the smartphone's current GPS location. In this paper, we investigated how such country-centric apps with location-based services can be employed to study the Chinese diaspora. Our methodology combines GPS hacking, automated task tools for mobile phones, and OCR to generate migration statistics for diaspora.  

As a case study, we applied our methodology to WeChat, an enormously popular app within China and among ethnic Chinese worldwide. Using WeChat, we collect data about the Chinese diaspora in 32 cities. 
The combined data provides interesting insights to the modern Chinese diaspora and how it has changed in recent years. To obtain a more complete picture of the Chinese diaspora, a larger number of cities would need to be investigated. 
We hope the data collection and analysis methodology described in this paper will help diaspora researchers to better understand immigration flows in our increasingly dynamic and fluid world.

\bibliographystyle{IEEEtran}
\bibliography{sigproc} 

\begin{thebibliography}{10}
\providecommand{\url}[1]{#1}
\csname url@samestyle\endcsname
\providecommand{\newblock}{\relax}
\providecommand{\bibinfo}[2]{#2}
\providecommand{\BIBentrySTDinterwordspacing}{\spaceskip=0pt\relax}
\providecommand{\BIBentryALTinterwordstretchfactor}{4}
\providecommand{\BIBentryALTinterwordspacing}{\spaceskip=\fontdimen2\font plus
\BIBentryALTinterwordstretchfactor\fontdimen3\font minus
  \fontdimen4\font\relax}
\providecommand{\BIBforeignlanguage}[2]{{%
\expandafter\ifx\csname l@#1\endcsname\relax
\typeout{** WARNING: IEEEtran.bst: No hyphenation pattern has been}%
\typeout{** loaded for the language `#1'. Using the pattern for}%
\typeout{** the default language instead.}%
\else
\language=\csname l@#1\endcsname
\fi
#2}}
\providecommand{\BIBdecl}{\relax}
\BIBdecl

\bibitem{belai2007enabling}
B.~H. Belai, ``{Enabling Diaspora Engagement in {Africa}: Resources,
  Mechanisms, and Gaps. Case study: {Ethiopia}},'' \emph{Ottawa: Association
  for Higher Education and Development}, 2007.

\bibitem{bloch2005development}
A.~Bloch, \emph{{The Development Potential of {Zimbabweans} in the Diaspora: A
  Survey of {Zimbabweans} Living in the {UK} and {South Africa}}}, ser.
  17.\hskip 1em plus 0.5em minus 0.4em\relax London: International Organization
  for Migration (IOM), 2005.

\bibitem{nworah2005study}
U.~Nworah, ``{Study on {Nigerian} Diaspora},'' \emph{Global Politician}, 2005.

\bibitem{zagheni2012you}
E.~Zagheni and I.~Weber, ``{You are Where You E-mail: Using E-mail Data to
  Estimate International Migration Rates},'' in \emph{Proceedings of the 4th
  Annual ACM Web Science Conference}.\hskip 1em plus 0.5em minus 0.4em\relax
  ACM, 2012, pp. 348--351.

\bibitem{weber2013studying}
B.~State, I.~Weber, and E.~Zagheni, ``{Studying Inter-national Mobility through
  {IP} Geolocation},'' in \emph{Proceedings of the sixth ACM International
  Conference on Web Search and Data Mining}.\hskip 1em plus 0.5em minus
  0.4em\relax ACM, 2013, pp. 265--274.

\bibitem{zagheni2014inferring}
E.~Zagheni, V.~R.~K. Garimella, I.~Weber \emph{et~al.}, ``{Inferring
  International and Internal Migration Patterns from {Twitter} Data},'' in
  \emph{Proceedings of the 23rd International Conference on World Wide
  Web}.\hskip 1em plus 0.5em minus 0.4em\relax ACM, 2014, pp. 439--444.

\bibitem{hawelka2014geo}
B.~Hawelka, I.~Sitko, E.~Beinat, S.~Sobolevsky, P.~Kazakopoulos, and C.~Ratti,
  ``{Geo-located {Twitter} as Proxy for Global Mobility Patterns},''
  \emph{Cartography and Geographic Information Science}, vol.~41, no.~3, pp.
  260--271, 2014.

\bibitem{messias2016migration}
J.~Messias, F.~Benevenuto, I.~Weber, and E.~Zagheni, ``{From Migration
  Corridors to Clusters: The Value of Google+ Data for Migration Studies},'' in
  \emph{Advances in Social Networks Analysis and Mining (ASONAM), 2016 IEEE/ACM
  International Conference on}.\hskip 1em plus 0.5em minus 0.4em\relax IEEE,
  2016, pp. 421--428.

\bibitem{xue2015thwarting}
M.~Xue, Y.~Liu, K.~Ross, and H.~Qian, ``{Thwarting Privacy protection on
  Location-based Social Discovery Services},'' \emph{Wiley Security and
  Communication Networks Journal}, 2015.

\bibitem{ding2014stalking}
Y.~Ding, S.~T. Peddinti, and K.~W. Ross, ``{Stalking Beijing from Timbuktu: A
  Generic Measurement Approach for Exploiting Location-based Social
  Discovery},'' in \emph{Proceedings of the 4th ACM Workshop on Security and
  Privacy in Smartphones \& Mobile Devices}.\hskip 1em plus 0.5em minus
  0.4em\relax ACM, 2014, pp. 75--80.

\bibitem{wang2016defending}
G.~Wang, B.~Wang, T.~Wang, A.~Nika, H.~Zheng, and B.~Y. Zhao, ``{Defending
  against {Sybil} Devices in Crowdsourced Mapping Services},'' in
  \emph{Proceedings of the 14th Annual International Conference on Mobile
  Systems, Applications, and Services}.\hskip 1em plus 0.5em minus 0.4em\relax
  ACM, 2016, pp. 179--191.

\bibitem{fin24tech}
\BIBentryALTinterwordspacing
Fin24, ``{China's WeChat Takes on WhatsApp in Africa},'' 2016. [Online].
  Available:
  \url{http://www.fin24.com/Tech/Mobile/chinas-wechat-takes-on-whatsapp-in-africa-20160724}
\BIBentrySTDinterwordspacing

\bibitem{techasia}
\BIBentryALTinterwordspacing
M.~Rao, ``{Who will be the WeChat of India? And What will it Do to Flipkart and
  Snapdeal?}'' 2016. [Online]. Available:
  \url{https://www.techinasia.com/wechat-india-flipkart-snapdeal}
\BIBentrySTDinterwordspacing

\bibitem{Forbes}
\BIBentryALTinterwordspacing
J.~Lim, ``{WeChat, One Of The World's Most Powerful Apps},'' 2014. [Online].
  Available:
  \url{http://www.forbes.com/sites/jlim/2014/05/19/wechat-one-of-the-worlds-most-powerful-apps}
\BIBentrySTDinterwordspacing

\bibitem{techasia2}
\BIBentryALTinterwordspacing
S.~Millward, ``{WeChat's global expansion has been a disaster},'' 2016.
  [Online]. Available:
  \url{https://www.techinasia.com/wechat-global-expansion-fail}
\BIBentrySTDinterwordspacing

\bibitem{similarweb}
\BIBentryALTinterwordspacing
J.~Schwartz, ``{The Most Popular Messaging App in Every Country},'' 2016.
  [Online]. Available:
  \url{https://www.similarweb.com/blog/worldwide-messaging-apps}
\BIBentrySTDinterwordspacing

\bibitem{PratoBook}
L.~Baldassar, G.~Johanson, N.~McAuliffe, and M.~Bressan, \emph{{Chinese
  Migration to Europe: Prato, Italy, and Beyond}}.\hskip 1em plus 0.5em minus
  0.4em\relax Springer, 2015.

\bibitem{WikiItaly}
\BIBentryALTinterwordspacing
{Wikipedia}, ``{Chinese people in Italy --- Wikipedia{,} The Free
  Encyclopedia},'' 2016. [Online]. Available:
  \url{https://en.wikipedia.org/wiki/Chinese_people_in_Italy}
\BIBentrySTDinterwordspacing

\bibitem{Pirepoint}
\BIBentryALTinterwordspacing
{Pacific Islands Report}, ``{China Is Biggest Source Of Foreign Investment,
  Tourists To Fiji},'' 2016. [Online]. Available:
  \url{http://www.pireport.org/articles/2016/11/29/china-biggest-source-foreign-investment-tourists-fiji}
\BIBentrySTDinterwordspacing

\bibitem{AntwerpChinatown}
\BIBentryALTinterwordspacing
{Chinatownology}, ``{Antwerp Chinatown},'' 2015. [Online]. Available:
  \url{http://www.chinatownology.com/chinatown_antwerp.html}
\BIBentrySTDinterwordspacing

\bibitem{Portugal}
\BIBentryALTinterwordspacing
{Algarve Daily News (www.algarvedailynews.com)}, ``{Chinese immigration highest
  in Portugal},'' 2014. [Online]. Available:
  \url{http://algarvedailynews.com/news/2702-chinese-immigration-highest-in-portugal}
\BIBentrySTDinterwordspacing

\bibitem{2010UScensus}
\BIBentryALTinterwordspacing
{U.S. Census Bureau}, ``{2010 Census Interactive Population Search: CA -- San
  Francisco city},'' 2010. [Online]. Available:
  \url{http://www.census.gov/2010census/popmap/ipmtext.php}
\BIBentrySTDinterwordspacing

\bibitem{herdaugdelen2016social}
A.~Herda{\u{g}}delen, L.~Adamic, W.~Mason \emph{et~al.}, ``{The Social Ties of
  Immigrant Communities in the {United States}},'' in \emph{Proceedings of the
  8th ACM Conference on Web Science}.\hskip 1em plus 0.5em minus 0.4em\relax
  ACM, 2016, pp. 78--84.

\bibitem{gordon1964assimilation}
M.~M. Gordon, \emph{{Assimilation in {American} life: The Role of Race,
  Religion, and National Origins}}.\hskip 1em plus 0.5em minus 0.4em\relax
  Oxford University Press, 1964.

\end{thebibliography}

\end{document}